\documentclass[%
 reprint,
 amsmath,amssymb,
 aps,
pra,
]{revtex4-2}
\usepackage{array}
\usepackage{graphicx}
\usepackage{bm}
\usepackage{bbm}
\usepackage{bbold}
\usepackage{amsmath}
\usepackage{braket}
\usepackage{color}
\usepackage{ulem}
\usepackage{tabularx}

\begin{document}

\title{Cavity Mode Initialization via a Rabi Driven Qubit}

\author{N. Karaev}
\email[Contact author: ]{natan.karaev@campus.technion.ac.il}
\author{E. Blumenthal}
\author{G. Moshel}
\author{A. A. Diringer}
\author{S. Hacohen-Gourgy} 
\affiliation{Department of Physics, Technion - Israel Institute of Technology, Haifa 32000, Israel}


\begin{abstract}
Microwave cavity modes with long coherence times are used in many different quantum computing systems. During normal operation of such systems, these modes, called memory modes, often need to be set to different coherent photonic occupations. In this work we present a novel technique we call Rabi Driven Reset in which the state of a memory mode is transferred into a decaying mode. This is done through a Rabi driven qubit which is coupled to both modes via sideband driving tones. The outcome of the method is the initialization of the memory mode at any required coherent state. Simulations are presented to demonstrate the effectiveness of this technique, along with a comparison to an existing coupling method. Our simulations predict an improvement of an order of magnitude in initialization times compared to existing methods.
\end{abstract}

\maketitle
\section{Introduction}

Quantum state initialization is a fundamental requirement for quantum technologies, both in sensing and in quantum information processing~\cite{Nielsen2012QuantumInformation}. A trivial quantum state initialization (or reset) method is to let the system decay to some steady state over time. However, as quantum systems are engineered to reach longer lifetimes, the reset time grows longer as well. As such, experiment repetition rates are limited by reset times which are orders of magnitude longer than coherent gate times. For example, in state of the art circuit quantum electrodynamics (circuit QED) systems, the operation times are tens of nanoseconds~\cite{Eickbusch2022FastQubit} whereas the decay times are hundreds of microseconds~\cite{Copetudo2024ShapingCQED}. Thus, the task of fast and efficient quantum state initialization proves to be a significant challenge which must be overcome.

Regardless of the physical platform, achieving reset that is faster than the decay time can only be realized through increased dissipation that can be turned on or off at will. This can be done either by measurement and feedback or by strongly coupling the system to a cold bath~\cite{Schliesser2008Resolved-sidebandOscillator,Peik1999SidebandTraps,Hamann1998Resolved-SidebandLattice,Leibrandt2009CavityIon,Meyer2019Resolved-SidebandNoise}. The two methods are seemingly different, but were shown to be equivalent~\cite{Cruikshank2017TheProtocols}.

Currently one of the most promising schemes for quantum information encoding is continuous variable encoding~\cite{Joshi2021QuantumQED}. Specifically in circuit QED, continuous variable encoding has seen significant growth in recent years~\cite{Campagne-Ibarcq2020QuantumOscillator,Lescanne2020ExponentialOscillator}.
In this platform, a transmon qubit is typically used as an ancilla for the control and readout of a long lived mode, called a memory mode, that stores the quantum state. Two reset methods for the memory mode have previously been employed~\cite{Sivak2023Real-timeBreak-even,Pfaff2017ControlledMemory,Milul2023SuperconductingTime}. In the first method, shown in Ref.~\cite{Sivak2023Real-timeBreak-even}, a memory mode excitation is transferred to the qubit which is subsequently reset using a measurement feedback protocol~\cite{Riste2012FeedbackMeasurement}. This cycle is repeated until the memory mode reaches a state sufficiently close to vacuum. Unfortunately, due to measurement and feedback time overheads, this method becomes impractical for cooling large photon numbers. In the second method, shown in Refs.~\cite{Pfaff2017ControlledMemory,Milul2023SuperconductingTime} the memory mode was coupled to a short lived mode, called a readout mode, using a pump tone which utilized the nonlinearity of an ancillary transmon. This method is seemingly suitable for cooling larger photon numbers, but in practice can only reach low cooling rates as it is limited by the anharmonicity of the transmon. 

In the current work we present an autonomous dissipative cooling method for the memory mode in a continuous variable circuit QED setup, that can surpass the cooling rates of previous methods. By Rabi driving the qubit and coupling it to both a memory mode and a readout mode using sideband drive tones, we are able to create significant coupling between the modes. The Rabi driven qubit is at an advantage compared to an undriven qubit since its frequency is much smaller than a qubit's typical frequency. This allows for the use of smaller detunings, which in turn increases the efficiency of the process. A similar sideband cooling method was previously implemented to cool down a transmon qubit~\cite{Murch2012Cavity-AssistedEngineering}, and here we expand it to cool a memory mode as well. This sideband cooling initializes the memory mode in a coherent state. To prepare the vacuum state a displacement with negligible duration can be applied.

\section{Theory}\label{sec:MemModCool}

\begin{figure*}
    \centering
        \includegraphics[width=1\linewidth]{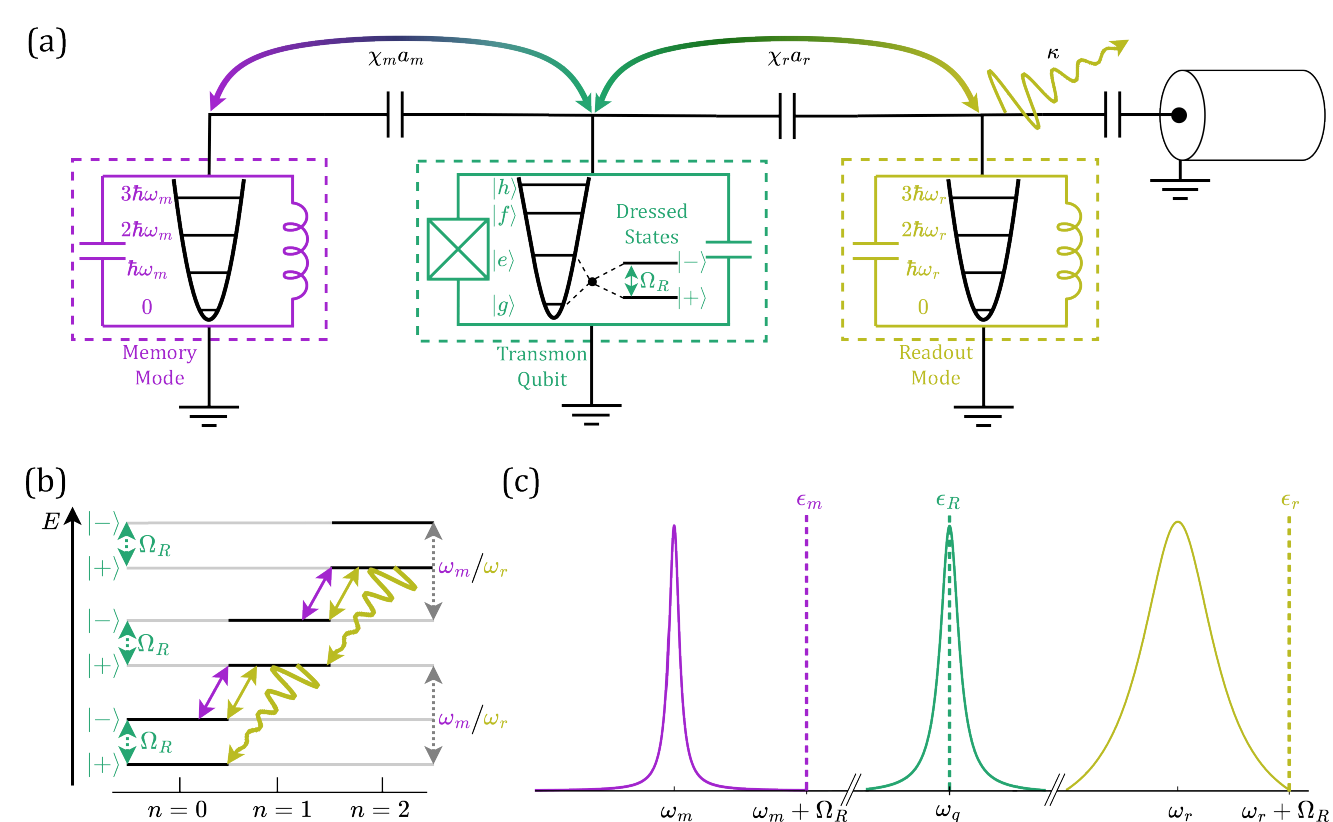}

\caption{System schematic diagrams. (a) Circuit Diagram. The memory mode (purple) and readout mode (yellow) are capacitively coupled to the transmon (green). The readout mode is also coupled to a transmission line which acts as a thermal bath. (b) Energy level diagram of the qubit-cavity system, for either the memory (purple) or readout (yellow) mode. The energy is presented in the lab frame. Bidirectional lines represent the cooling sideband drives, while decaying lines represent spontaneous decay in the cavity. (c) Frequency response of the different components.  Dashed lines represent drives, wherein $\epsilon_m$ and $\epsilon_r$ are cavity drives and $\epsilon_R$ is the Rabi drive.}
\label{fig:system}
\end{figure*}

The initialization scheme we propose is presented in Fig.~\ref{fig:system}. It is an expansion on the qubit sideband cooling method of Ref.~\cite{Murch2012Cavity-AssistedEngineering} (see Appendix~\ref{supp:SidebandCooling}), wherein we consider a memory mode and a readout mode that are dispersively coupled to a Rabi driven transmon.  We treat the transmon as a qubit, assuming that the Rabi frequency $\Omega_R$ is much smaller than the transmon anharmonicity. The eigenstates of the qubit are thus the $\ket{-}$ and $\ket{+}$ states, with energy difference $\Omega_R$. We apply sideband drives to both modes, with a detuning that is equal to the Rabi frequency. 
We use a Rabi driven qubit instead of an undriven qubit because a standard qubit's energy is of order of a few GHz, which would make the sideband drives significantly detuned from the cavities' frequencies. The density of states of the cavities at detunings of a few GHz are negligible (unlike for a Rabi driven qubit, as in Fig.~\ref{fig:system}(c)) which would make the following process extremely inefficient (as shown in Ref.~\cite{Murch2012Cavity-AssistedEngineering}). Thus, the use of a Rabi driven qubit is necessary. 

In a frame rotating with the two sideband drives and the Rabi drive, the Hamiltonian is
\begin{equation}
    \begin{split}
        H=-\Omega_Ra_r^\dagger a_r-\Omega_Ra_m^\dagger a_m-\frac{\Delta_q+\chi_r+\chi_m}{2}
        \sigma_z\\-\frac{\Omega_R}{2}\sigma_x-\chi_r a_r^\dagger a_r\sigma_z-\chi_m a_m^\dagger a_m\sigma_z\\+\epsilon_r(a_r^\dagger+a_r)+\epsilon_m(a_m^\dagger+a_m),
    \end{split}
    \label{eq:convHamilt}
\end{equation}
where $\Delta_q$ is the detuning of the Rabi drive, $a_{m,r}$ are field annihilation operators, $\chi_{m,r}$ are dispersive coupling strengths with the transmon, and $\epsilon_{m,r}$ are the complex drive amplitudes. The $r$ and $m$ subscripts refer to the readout and memory modes, respectively. 

Under constant drives, the cavities' steady states are coherent states with amplitudes $\bar{a}_m$ and $\bar{a}_r$. Calculating these amplitudes in the absence of a qubit we get

\begin{equation}
        \bar{a}_m=\frac{\epsilon_m}{\Omega_R}~, ~\bar{a}_r=\frac{\epsilon_r}{\Omega_R+\frac{i\kappa}{2}},
    \label{eq:cavOccupation}
\end{equation}
where $\kappa$ is the decay rate of the readout mode and we assume that the memory mode natural decay rate is negligible. We require $\kappa$ to be smaller than $\Omega_R$ to make sure only a single sideband is excited by our drives. Our method will initialize the system's cavity modes at these states. We move to a displaced frame by eliminating the drive terms and substituting $a_{m,r}\rightarrow \tilde{a}_{m,r}+\bar{a}_{m,r}$. By setting $\Delta_q=-\chi_m(|\bar{a}_{m}|^2+1)-\chi_r(|\bar{a}_{r}|^2+1)$, we eliminate the constant Stark shift term. Now we rename $\sigma_x\leftrightarrow\sigma_z$ and go to the frame rotating at the Rabi frequency. After applying the rotating wave approximation (RWA) to neglect terms that oscillate at $\Omega_R$, we are left with

\begin{equation}
    \begin{split}
        H=(g_r^\star \tilde{a}_r\sigma_++g_r\tilde{a}_r^\dagger\sigma_-)\\+(g_m^\star \tilde{a}_m\sigma_++g_m\tilde{a}_m^\dagger\sigma_-),
        \end{split}
    \label{eq:convHamiltDisplacedRWAZero}
\end{equation}

where $g_{m,r}\equiv|\chi_{m,r}|\bar{a}_{m,r}$ is the coupling coefficient. This coefficient represents the transfer rate of excitations. This ``dual'' Jaynes-Cummings Hamiltonian transfers excitations between the memory mode and the qubit, and between the qubit and the readout mode. See Appendix~\ref{supp:RotatingWave} for more detailed explanations.

We demand that the excitations of the readout mode decay to the transmission line before they interact with the qubit, so that the readout mode can be regarded as a Markovian bath. This is achieved by setting
\begin{equation}
    \begin{split}
        |\chi_r\bar{a}_r|\leq\frac{\kappa}{2}.
    \end{split}
    \label{eq:weakCoupling}
\end{equation}
Since the qubit can not contain more than one excitation, this bound on the coupling strength ensures that the readout mode is excited at a slower rate than it decays. This limited excitation rate leads to a limit on the rate of photon loss from the readout mode to the transmission line. As the Hamiltonian preserves the number of excitations, this is also the limit of the cooling rate of the memory mode. We find that the maximum cooling rate is (see Appendix~\ref{supp:MaxCoolingRate})
\begin{equation}\label{eq:MaxCoolingRate}
    |\dot{\tilde{n}}_m|^\mathrm{max}=\frac{\kappa}{4},
\end{equation}
where $\tilde{n}_m$ is the number of photons in the memory mode in the displaced frame.

It is important to note that the cooling of any system through a lossy mode with a finite decay rate is limited by non-Markovian effects. Usually, these effects will lead to oscillations in the cooling rate of the system. For example, if we were to create a strong direct coupling between the memory mode and the readout mode we would see excitations oscillating back and forth between them throughout the cooling process. In our case, the memory mode is coupled to the qubit, and it undergoes a Rabi oscillation at a rate of $n_m\chi\bar{a}_m$. However, even when the memory mode is highly populated, the qubit acts as a bottleneck that sets the cooling rate of the readout mode to be in the Markovian regime. This leads to a constant photon loss rate that saturates the bound of Eq.~\ref{eq:MaxCoolingRate}. This bound saturation ceases to apply once the memory mode is cooled down to a few photons. In order to equalize the excitation transfer rates to and from the qubit, we match the effective coupling between each mode and the qubit by setting

\begin{equation}
    \begin{split}
        g_r=\chi_r|\bar{a}_r|=\chi_m|\bar{a}_m|=g_m.
    \end{split}
    \label{eq:eqRates}
\end{equation}

In what follows, we shall be refer to this cooling process as Rabi Driven Reset (RDR).

\section{Results}\label{sec:displacedHamilt}\label{sec:fullHamilt}
To demonstrate the dynamics of this method we performed several simulations using Python's  QuTiP package~\cite{Johansson2012QuTiP:Systems,Johansson2013QuTiPSystems}. In order to compare our method to the cooling rates of a state of the art system, the parameters of the memory mode, readout mode, and qubit are mostly taken to be the same as in Ref.~\cite{Milul2023SuperconductingTime}, in which the initialization method shown in Ref.~\cite{Pfaff2017ControlledMemory} was used in a system with cutting edge cavity coherence times. Specifically $\kappa/2\pi=0.419$ MHz, $\Omega_R/2\pi=20$ MHz, $\chi_r/2\pi=-1.3$ MHz, and $\chi_m/2\pi=-0.042$ MHz. The dephasing and energy relaxation of the memory mode are ignored as they should not interfere with the cooling. The qubit's dephasing and energy relaxation rates are also ignored for simplicity's sake, though simulations of their effect (when they are not negligible) can be found in Appendix~\ref{supp:QubitParams} as part of an analysis of simulations with different parameters than those noted above. Additional technical details regarding the simulations are shown in Appendix~\ref{supp:Sim}.

\subsection{The full and displaced Hamiltonians}\label{sec:fullHamiltOnly}
\begin{figure}
    \centering
    \includegraphics[width=0.95\linewidth,trim = 0 30 25 13,clip]{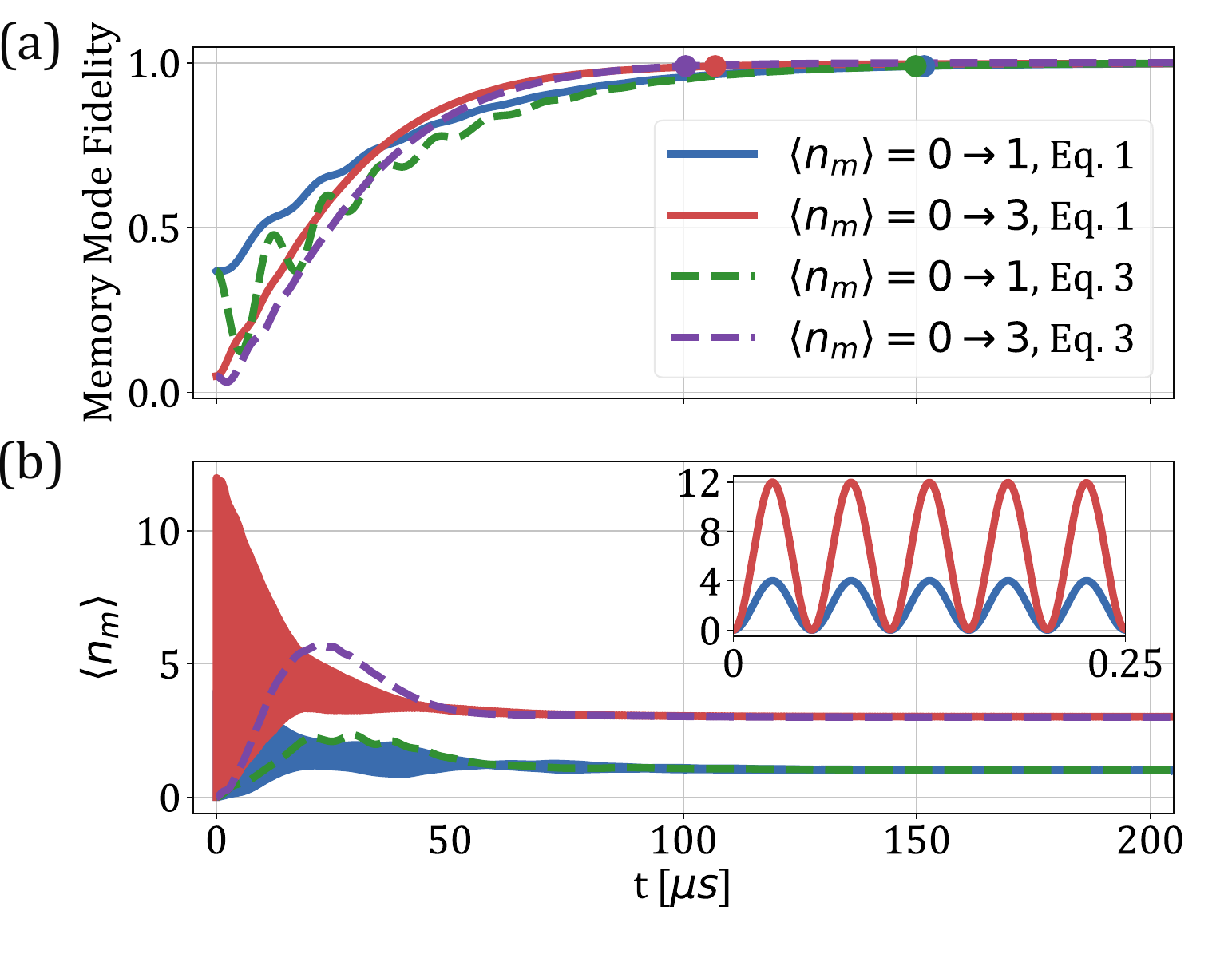}
    \caption{Simulations of the Hamiltonian in Eq.~\ref{eq:convHamilt} (solid lines) and Eq.~\ref{eq:convHamiltDisplacedRWAZero} (dashed lines). Photonic occupations are $\left\langle n_m\right\rangle=0\rightarrow1$ (blue and green) and $\left\langle n_m\right\rangle=0\rightarrow3$ (red and purple). (a) Time evolution of the memory mode fidelity with respect to its required final state. The dots mark $0.99$ fidelity. Simulations to higher photonic occupations reach 0.99 fidelity earlier. (b) Expectation value of the photonic occupation of the memory mode. The vertical thickness of the blue and red graphs are due to high order oscillations. A close up of the oscillations is shown in the inset.}
    \label{fig:fullHamiltSim}\label{fig:maxRate}
\end{figure}

We will show some fundamental simulations of our method. The results of these simulations are shown in Fig.~\ref{fig:maxRate}.
We first simulate the full Hamiltonian presented in Eq.~\ref{eq:convHamilt} to show the viability of this method. Since we have designed this Hamiltonian to take the system from any state to some predetermined coherent state, we first simulate with vacuum initial states in both modes. The drives are set according to Eq.~\ref{eq:cavOccupation} and Eq.~\ref{eq:eqRates} so that the final photonic occupations are $\left\langle n_r\right\rangle=1.0\cdot10^{-3}$ and $\left\langle n_m\right\rangle=1.0$ which obey Eq.~\ref{eq:eqRates} and Eq.~\ref{eq:weakCoupling}. This sets the drive amplitudes as $\epsilon_r=4.1$ MHz and $\epsilon_m=125.7$ MHz. It is simplest to assess the quality of the results of the system by calculating the fidelity of the memory mode with its required coherent final state, i.e. the expected probability of finding the memory mode in state $\ket{\alpha=\bar{a}_m}$. Even for these relatively small photonic occupations, the RDR process sets the required states an order of magnitude faster than the cooling times of the method in Ref.~\cite{Milul2023SuperconductingTime}.

We then simulate our method for higher final photonic occupations of $\left\langle n_r\right\rangle=3.0\cdot10^{-3}$ and $\left\langle n_m\right\rangle=3.0$, which correspond to drive amplitudes of $\epsilon_r=7.1$ MHz and $\epsilon_m=217.7$ MHz. It is clear that the increase in photonic occupations and drive amplitudes has reduced the amount of time required to reach the final state. In both of these simulations the oscillations of $n_m$ are caused by the sideband detuning, and thus they oscillate at the detuning frequency.

In order to reach higher photonic occupations, and thus the rate cap of the RDR method, we will need to simulate higher photonic occupations. In order to perform these simulations we will use the displaced Hamiltonian of Eq.~\ref{eq:convHamiltDisplacedRWAZero}. We have thus performed both of the previously mentioned simulations again but for the displaced Hamiltonian. It is possible to see that the simulation reaches comparable results for both of the final photonic occupations. The oscillations in $n_m$ do not appear in the simulations of the displaced Hamiltonian as the terms which cause them are not present in Eq.~\ref{eq:convHamiltDisplacedRWAZero}. However the closeness of the overall simulation and performance of the displaced Hamiltonian to the full Hamiltonian of Eq.~\ref{eq:convHamilt} allow us to go forward and simulate the maximum cooling rate achievable with the RDR method at high photonic occupations.

\subsection{Cooling rate}\label{sec:maxRate}

As shown in section~\ref{sec:MemModCool}, an inherent feature of our method is that the qubit acts as a bottleneck which holds, at most, a single excitation at a given time. Thus, even when the long lived memory mode has a high occupation, it is still being cooled at a constant rate, until $\tilde{n}_m\approx1$. We will now compare the performance of the cooling of the method in Ref.~\cite{Milul2023SuperconductingTime} with the cooling we can achieve with our method. 

In order to extract the maximum constant cooling rate, we simulate the cooling using Eq.~\ref{eq:convHamiltDisplacedRWAZero} from an initial photonic occupation of $\left\langle n_m\right\rangle=99.4$ to a final photonic occupation of $\left\langle n_m\right\rangle=24.8$ in the non-displaced frame, which is translated to a change of the occupation from $\left\langle \tilde{n}_m\right\rangle=24.8$ to $\left\langle \tilde{n}_m\right\rangle=0$ in the displaced frame. These final occupation numbers are not arbitrary but are the limits of weak coupling of Eq.~\ref{eq:weakCoupling}, and we choose the starting photonic occupation to simply have double the annihilation operator expectation value of the final occupation. The drive values are thus $\epsilon_r=20.2$ MHz and $\epsilon_m=626.6$ MHz. For this section and all subsequent sections, the readout mode's initial occupation is set according to Eq.~\ref{eq:eqRates}, to start with some initial occupation.

We extract the maximum change rate from the constant region of the displaced photonic occupation. The rate in this region is $d\left\langle \tilde{n}_m\right\rangle/dt = 0.641\  1/\mu s$, whereas the maximum cooling rate expressed in section~\ref{sec:MemModCool} is $\kappa/4=0.658\ 1/\mu s$. We can see that these values are quite close. It is tempting to consider increasing the drive in order to accelerate the rate, but as we will see in section~\ref{sec:differentDrives} the benefit is quite small. This is because any increase in the drive brings us out of the weak coupling regime of Eq.~\ref{eq:weakCoupling} and causes oscillations of excitations between the readout mode and the qubit, which causes the decay of the excitations to be less effective.
 
The theoretical basis of the method in Ref.~\cite{Milul2023SuperconductingTime} is explained in depth there, but we will list some basic facts regarding it to assist with the comparison of the two methods. It contains the same basic components as in our method: a memory mode, a qubit, and a readout mode. The activatable  $\kappa$ applied to the memory mode is directly proportional to $\xi$, which is the displacement of the qubit. Naively, this means it is possible to cool at any rate by applying increasingly large displacements of the qubit. However, an implicit limitation of the method is that it requires $\xi\ll1$. For our calculation of the reported cooling method performed in Ref.~\cite{Milul2023SuperconductingTime}, we have calculated an exponential decay. This decay is characterized by a decay time $T=0.6 ms$, which is caused by a displacement $\xi=0.15$. Most similar experimental setups will not be able to increase $\xi$ much more than the above value. Seeing that we have used the parameters reported in Ref.~\cite{Milul2023SuperconductingTime}, our calculated exponential decay is representative of the cooling performance attainable with their method, and can be directly compared to the simulated performance of our method.

\begin{figure}[]
    \centering
    \includegraphics[width=0.95\linewidth,trim = 20 30 28 25,clip]{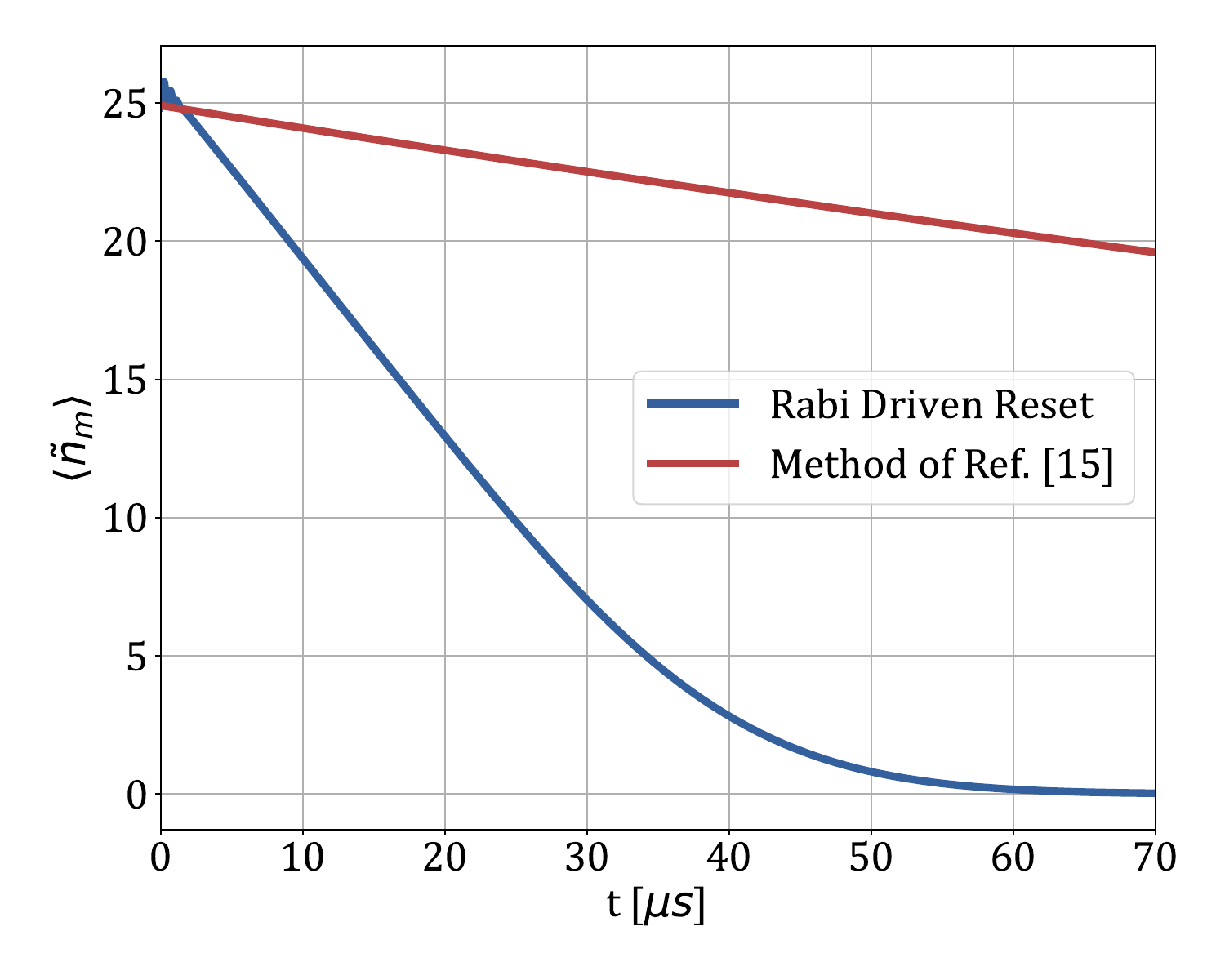}
    \caption{Displaced photonic occupation for RDR and for the method in Ref.~\cite{Milul2023SuperconductingTime} in blue and red, respectively. The RDR simulation is performed using Eq.~\ref{eq:convHamiltDisplacedRWAZero}. The red graph is an exponential decay calculated with the parameters in Ref.~\cite{Milul2023SuperconductingTime} and transformed to the displaced frame.}
    \label{fig:WizComp}
\end{figure}

The simulation of our method is shown in Fig.~\ref{fig:WizComp}, along with the exponential cooling calculated from Ref.~\cite{Milul2023SuperconductingTime}. The advantage of our method is clear: It provides a fast constant cooling rate which, under a certain photonic occupation (and obeying the limitation of $\xi$), will always outperform the method in Ref.~\cite{Milul2023SuperconductingTime}. 

\begin{figure*}[t]
\centering    \includegraphics[width=0.95\linewidth,trim = 100 45 100 50,clip]{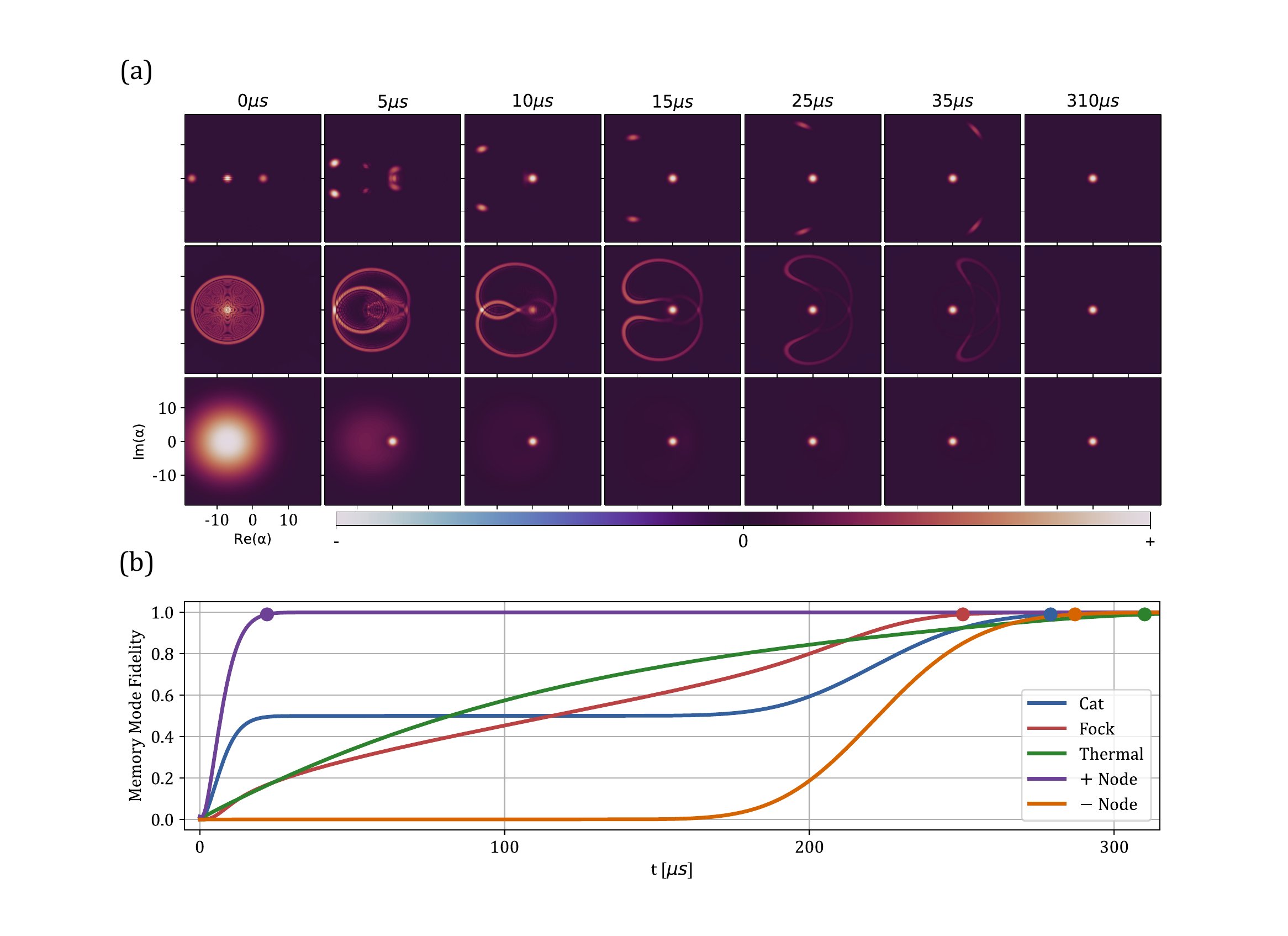}
\caption{Cooling of different initial states in the displaced frame. The non-displaced photonic occupations for each state are $\left\langle n_m\right\rangle=49.6\rightarrow24.8$ (a) The Wigner function of the quantum state during the cooling. The top, middle and bottom rows represent an initial Cat state, Fock state and Thermal state, respectively. Each individual graph is normalized for maximum contrast, thus the colorbar is qualitative. (b) The fidelity values of the simulations of cooling from different initial states to the required final coherent state using the Hamiltonian of Eq.~\ref{eq:convHamiltDisplacedRWAZero}. Blue, red and green data represent initial Cat, Fock and Thermal states. Purple and orange data represent initial states containing the positive and negative coherent nodes of the cat state, respectively, after appropriate normalization. The dots are the times at which the fidelity value is $0.99$ at times $t=279\mu s$, $t=250\mu s$ and $t=310\mu s$ for Cat, Fock and Thermal initial states, respectively. The positive and negative coherent nodes arrive at $0.99$ fidelity values for $t=22\mu s$ and $t=287\mu s$, respectively.}
\label{fig:InitCompare}
\label{fig:Movie}

\end{figure*}

\subsection{Cooling of different initial states}
In order to further demonstrate the efficiency of our method we simulate a cooling process to the same final state as in section~\ref{sec:maxRate} from different initial states: a cat state (defined as a superposition of two coherent states $\frac{1}{\mathcal{N}}(\ket{-\alpha}+\ket{\alpha})$, wherein $\mathcal{N}$ is a normalization factor), a fock state (defined as a state with a specific number of photons $\ket{n}$), and a thermal state. Additionally, in order to better analyze the evolution of the cat state, we performed two additional simulations using the positive and negative coherent parts of the cat state, dubbed ``$+$ Node'' and ``$-$ Node'', respectively. We chose all of these states to have the same final photonic occupation $n_m$ as in section~\ref{sec:maxRate}  ($\left\langle n_m\right\rangle=24.8$), and we initialized these states at a photonic occupation that is double the final occupation ($\left\langle n_m\right\rangle=49.6$). The fidelity results of the simulations and select Wigner function snapshots of the processes are shown in Fig.~\ref{fig:Movie}.

While it is clear that the initial state affects the arrival time at the final state, with the $+$ node being shifted significantly more quickly to the final coherent state as compared with the other initial states, all of them reach the required final state in $t\leq310\mu s$, as compared with several ms for the other methods. This is a clear demonstration as to the versatility of this method: It is effective regardless of the initial state of the cavity. Additionally, it is possible to observe that there is a clear correlation between the simulated behavior of the cat state and those of its constituent nodes: a clear shift in the fidelity rise occurs at the same time as when the $+$ node arrives at its final state, and the cat state's fidelity continues to increase with the fidelity increase of its $-$ node. This demonstrates a linear nature to this cooling method wherein the constituent parts of a state are transferred consistently both when they are separate and as part of a ``sum'' state.

\subsection{Drive power}\label{sec:differentDrives}

As mentioned previously, the main application of RDR is to perform state initialization. Specifically, when starting with a given arbitrary state, if we were to transfer it to some known coherent state then we would be able to near instantly (via displacement) shift it to some other coherent state. From Eq.~\ref{eq:eqRates} it is clear that, at least in the ideal theory, the excitation transfer rate increases with increased drive, as long as the weak coupling condition of Eq.~\ref{eq:weakCoupling} is obeyed. 

\begin{figure}[]
    \centering
    \includegraphics[width=0.95\linewidth,trim = 28 30 10 10,clip]{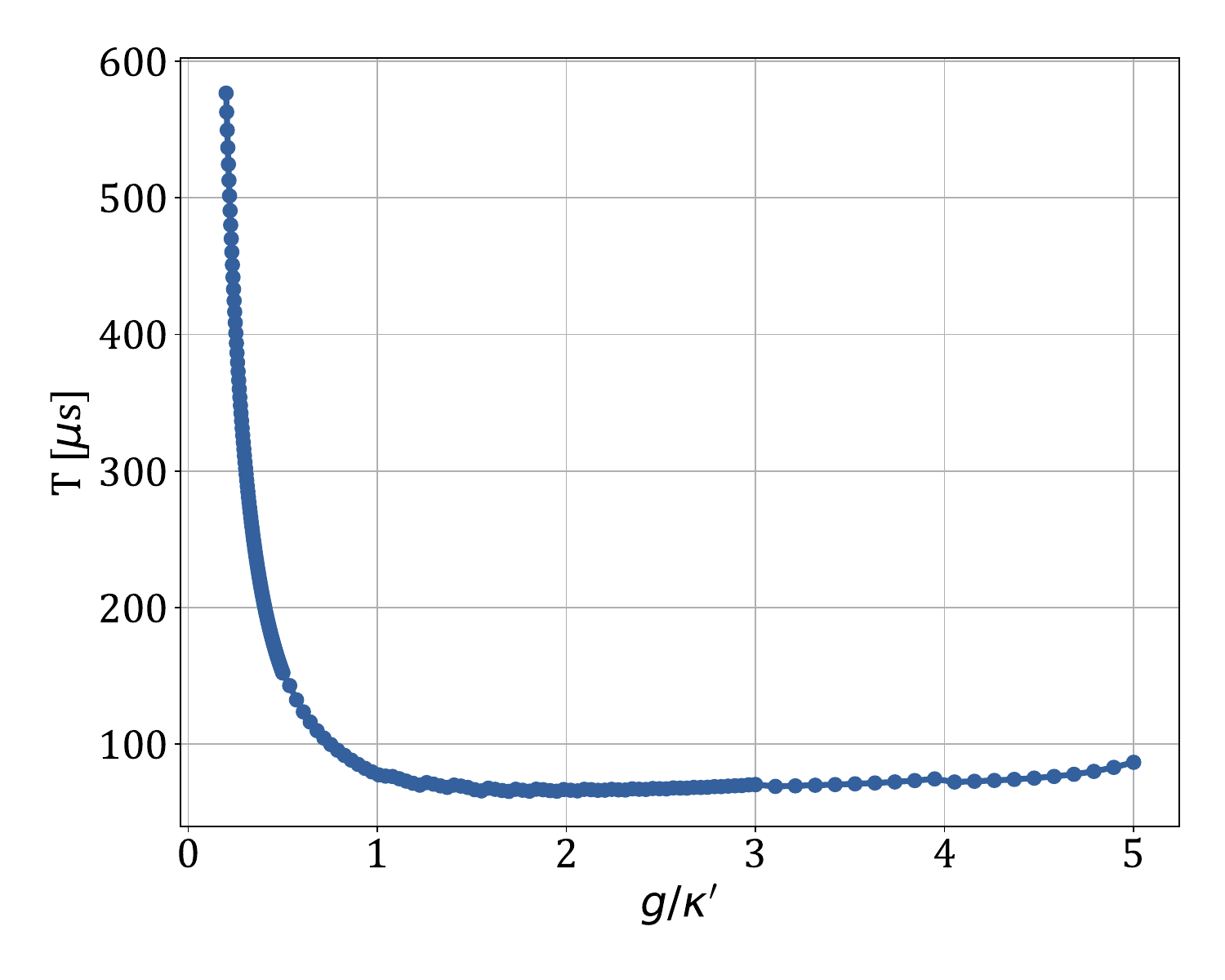}
    \caption{Cooling times as function of normalized drive. Initial state is a thermal state with photonic occupation of $\left\langle n_m\right\rangle=1$. Normalized drive $g/\kappa'=g_m/\kappa'=a_m\chi_m/\kappa'$ is varied by changing $a_m$, keeping all other parameters constant.}
    \label{fig:FidTimes}
\end{figure}

In order to test whether there is some extremum in cooling times as a function of drive power, we have performed simulations of cooling of a thermal state with an initial photonic occupation of $\left\langle n_m\right\rangle=1$ to different final coherent states, which also correspond to different drives according to Eq.~\ref{eq:cavOccupation}. The cooling time $T$ is defined as the time to reach a $0.99$ fidelity value of the required final state. In order to perform many simulations efficiently well past the weak coupling limit, we used a new decay rate $\kappa'=\kappa/10$. This reduced the final photonic occupations and allowed us to employ a smaller (and more computationally efficient) Hilbert space. The simulation results are displayed in Fig.~\ref{fig:FidTimes} as a function of normalized drive power $g/\kappa'$, wherein the limit for weak coupling presented in Eq.~\ref{eq:weakCoupling} is reached for $g/\kappa'=0.5$.

It is clear from the simulations that while increasing the drive power beyond weak coupling shortens $T$ slightly, there is less and less to gain with each increase, until at approximately $g/\kappa'=1$ the cooling times reach a plateau. With increases in normalized drive beyond $g/\kappa'=1$, the method starts to break down as its performance worsens. It is also interesting to note that higher drives beyond weak coupling displayed oscillatory effects in the cooling process, which seem to slow the cooling down. This is similar to the strong coupling effects described in Ref.~\cite{Murch2012Cavity-AssistedEngineering}, though less dramatic. Thus for this type of state initialization a relatively low power drive (compared to the power at $g/\kappa'=1$) is required to achieve cooling times which are comparable to the fastest possible times.

\section{Conclusion}
We have shown a novel microwave cavity coupling method. This method couples a long lived mode to a short lived mode via a Rabi driven qubit. We tested the  applicability of this method computationally and we found it to be stable and efficient, achieving faster cooling rates than other methods with experimentally viable parameters. This faster cooling rate is kept constant until photonic occupations of around $\tilde{n}_m\approx1$, and since many cutting edge bosonic codes employ states with $n_m>1$~\cite{Heeres2017ImplementingOscillator,Chou2018DeterministicQubits,Gao2019EntanglementInteraction,Xu2020DemonstrationQubits,Reinhold2020Error-correctedQubit}, this method is poised to be useful for bosonic quantum computation. The upper bound of the cooling rate of our method is set by both fundamental and experimental limitations. Fundamentally, it can not exceed the decay rate of the readout mode, which is usually around a few MHz. Experimentally, it is limited by the maximal effective coupling rate that can be generated while keeping the memory mode's long coherence times. The drive amplitude is practically limited to a few hundred MHz~\cite{Eickbusch2022FastQubit}, which in turn limits $a_m$ to be of an order of $10$, and the dispersive coupling strength is limited to tens of kHz to avoid unwanted self- and cross-Kerr effects. These two bounds set the cooling rate to about a MHz. In addition, for the RWA to be applicable, $\Omega_R$ must be higher than all the rest of the rates of the system. As $\Omega_R$ can realistically reach a few tens of MHz, this limit coincides with the previous ones.

We have thus demonstrated a very potent technique for cavity preparation via coupling to a cavity with significant decoherence. It has a range of potential applications as a tool for quantum computing systems, since it is in fact a versatile method to prepare coherent states in cavities, to empty cavities (via coherent state preparation followed by displacement) and transfer excitations by coupling a readout mode to a transmission line. This method could be enhanced even further by performing feedback and measurement when the state is close to its requires final state, as in Ref.~\cite{Sivak2023Real-timeBreak-even}. While future applications of this method could vary, it could serve as a basis for the fast preparation of high coherence cavities in quantum computing systems.

\begin{acknowledgments}
This research was supported by the Israeli Science Foundation (ISF), Pazi foundation, and Technion's Helen Diller Quantum Center.
\end{acknowledgments}

\appendix

\section{SIDEBAND COOLING OF A SUPERCONDUCTING QUBIT}\label{supp:SidebandCooling}
In order to understand our method, it is useful to describe a setup for sideband cooling of a supercondcting qubit. 
In such a setup a qubit of frequency $\omega_q$ and a microwave readout mode of frequency $\omega_c$ and decay rate $\kappa$ are dispersively coupled, with their frequencies far detuned from each other. The straightforward method one may employ to perform cooling is to directly drive a sideband by applying a drive detuned from the readout mode frequency by $\Delta=\omega_q-\omega_c$. In practice the drive would create a very weak interaction and lead to an ineffective cooling rate.

As described by Murch et al.~\cite{Murch2012Cavity-AssistedEngineering}, we can be significantly more efficient by Rabi driving the qubit and applying a sideband detuned by the Rabi frequency. Consider such a qubit driven perfectly at the energy difference of the $\ket{g}$ and $\ket{e}$ levels by a Rabi frequency $\Omega_R$. We neglect the $\ket{f}$ state and following states by assuming high enough anharmonicity and an appropriately weak Rabi drive. The new eigenstates are the dressed states $\ket{+}$ and $\ket{-}$, with an energy difference $\Omega_R$ between them. We now set the cavity drive detuning as $\Delta=\Omega_R$. Since this detuning is much closer to $\kappa$, we get a much stronger interaction for a detuned sideband cooling drive operating on the dressed states and the readout mode states. In the cavity and qubit drives frame of reference the Hamiltonian of such a system is 
\begin{equation}
    \begin{split}
        H=-\Omega_Ra^\dagger a-\frac{\Delta_q+\chi}{2}\sigma_z-\frac{\Omega_R}{2}\sigma_x\\-\chi a^\dagger a\sigma_z+\epsilon_c(a^\dagger+a),
    \end{split}
    \label{eq:qubitCoolingHamilt}
\end{equation}
where $a$ is the cavity annihilation operator, $\Delta_q$ is the Rabi drive's detuning, $\chi$ is the dispersive coupling strength, and $\epsilon_c$ is the cavity drive amplitude. 

The steady state occupation of the cavity for a given drive is given by the second expression in Eq.~\ref{eq:cavOccupation}. We displace $a$ by its steady state occupation $\bar{a}$ as $a=\bar{a}+\tilde{a}$ where $\tilde{a}$ is the displaced operator, after which $\Delta_q$ can be set as $\Delta_q=-\chi(\bar{a}^\star\bar{a}+1)$ to cancel all $\chi\sigma_z$ terms. A Hadamard rotation to exchange the $x$ and $z$ axes changes the Hamiltonian to the simpler form (ignoring rapidly rotating terms)
\begin{equation}
        H=-\Omega_R\tilde{a}^\dagger \tilde{a}-\frac{\Omega_R}{2}\sigma_z\\-\chi(\bar{a}^\star \tilde{a}\sigma_++\bar{a}\tilde{a}^\dagger\sigma_-).
    \label{eq:qubitCoolingHamiltDisplaced}
\end{equation}

This is the familiar form of the Jaynes-Cummings Hamiltonian, transferring excitations between a cavity mode and the qubit. It is possible to see that the coefficient of $a\sigma_+$ is the energy scale of the difference between the new eigenstates $\ket{n,\pm}\equiv(\ket{n,g}\mp\ket{n-1,e})$. Let us define this value as the coupling coefficient $g\equiv|\chi|\bar{a}$. 

For some arbitrary number of photons in the cavity $\ket{n-1}$, if the qubit is in the $\ket{e}$  state,the total state is a sum of the new eigenstates $\ket{n-1,e}=((\ket{n-1,e}-\ket{n,g})+(\ket{n-1,e}+\ket{n,g}))/2$. These new eigenstates gain phase at a rate of order $g$, which means after time of order $1/g$ the pure $\ket{g,n}$ state has become a pure $\ket{e,n-1}$ state. Without decay this process would continue to occur, as the qubit state changes from ground to excited and back again as the cavity photon count goes up and down by $1$. This means the coupling coefficient $g$ is the excitation exchange rate between the qubit and the cavity.

Adding in the decay rate, we assume that photons transfer qubit excitation from the higher energy dressed state to the lower energy dressed state while increasing the cavity mode photon number at a rate lower than its decay rate. This creates a two-step transfer process which stabilizes the qubit at the $\ket{+}$ dressed state. We can to see that choosing a negative detuning $\Delta=-\Omega_R$ instead causes stabilization at the $\ket{-}$ dressed state.

In essence, because we applied the drive to the cavity as previously described we have stabilized the system at a specific photonic occupation and dressed state. We remove any shift from these two specified parameters via the drive (increasing the photonic occupation) and the cavity's decay (decreasing the photonic occupation).
\section{FORMS OF THE HAMILTONIAN}\label{supp:RotatingWave}
\begin{figure}
    \centering
    \includegraphics[width=0.95\linewidth,trim = 0 30 20 10,clip]{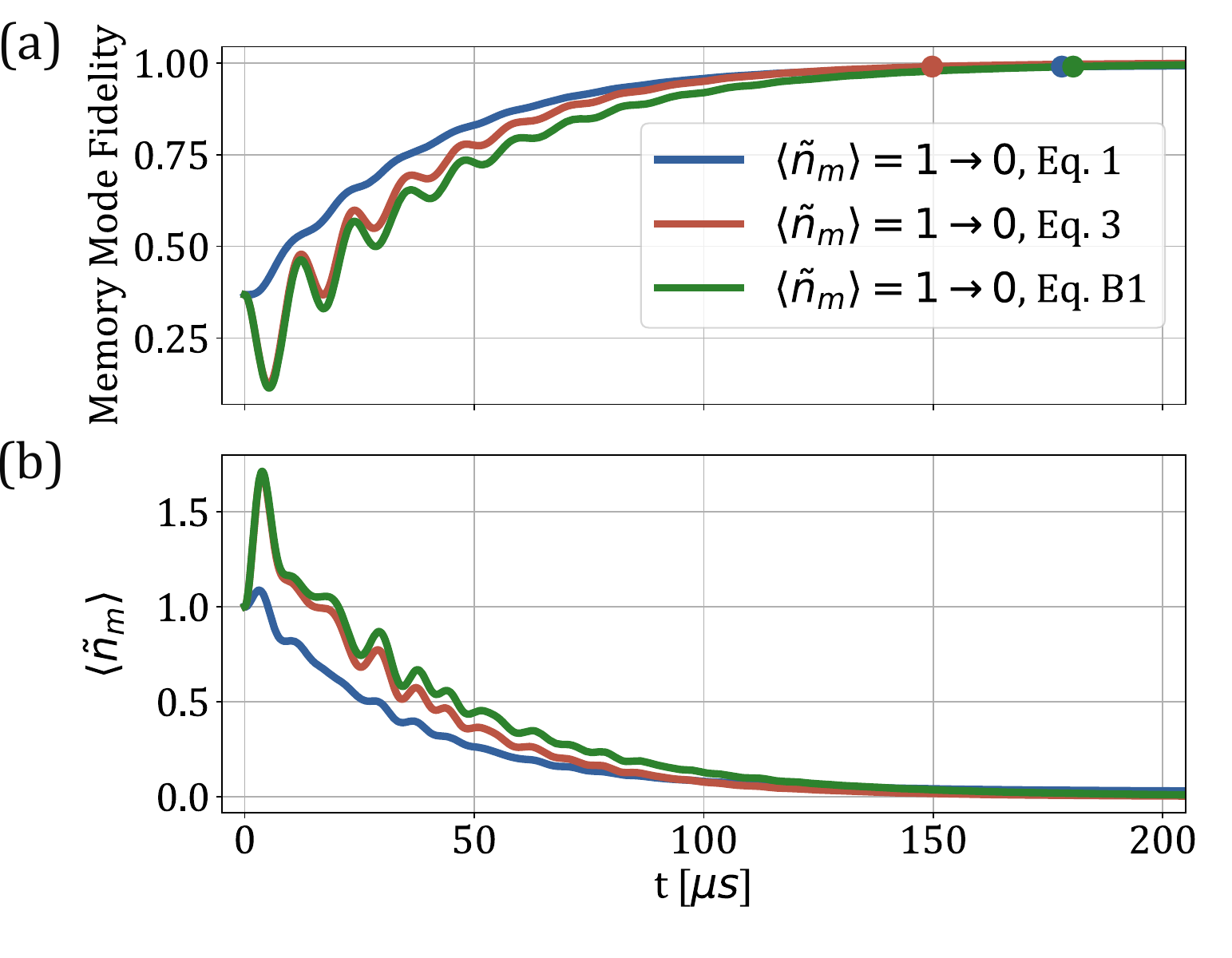}
    \caption{Simulation of the Hamiltonian in Eq.~\ref{eq:convHamilt} (blue), Eq.~\ref{eq:convHamiltDisplacedRWAZero} (red), and Eq.~\ref{eq:convHamiltDisplaced} (green) with the first simulation parameters of section~\ref{sec:fullHamiltOnly}. (a) The fidelity of the memory mode to its required displaced final state of $\ket{\alpha=0}$ in the displaced frame. The dots are the moments for each simulation that the fidelity reaches
a value of 0.99. (b) The displaced photonic occupation of the memory mode.}
    \label{fig:DisplacedHamiltSim}
\end{figure}

In order to test the validity of the final form of our Hamiltonian, shown in Eq.~\ref{eq:convHamiltDisplacedRWAZero}, we compare it to the initial form of the Hamiltonian in Eq.~\ref{eq:convHamilt}, and to

\begin{equation}
    \begin{split}
        H=-\Omega_R\tilde{a}_r^\dagger \tilde{a}_r-\Omega_R\tilde{a}_m^\dagger \tilde{a}_m-\frac{\Omega_R}{2}\sigma_z\\-\chi_r(\bar{a}_r^\star \tilde{a}_r+\bar{a}_r\tilde{a}_r^\dagger+\tilde{a}_r^\dagger \tilde{a}_r)\sigma_x\\-\chi_m(\bar{a}_m^\star \tilde{a}_m+\bar{a}_m\tilde{a}_m^\dagger+\tilde{a}_m^\dagger \tilde{a}_m)\sigma_x,
    \end{split}
    \label{eq:convHamiltDisplaced}
\end{equation}

which is similar to the final Hamiltonian, but before the application of the frame shifts to the Rabi drive and readout and memory mode drives, and before the rotating wave approximation.
In order to compare the Hamiltonians we have performed the first simulation described in section~\ref{sec:fullHamiltOnly} for all of them. The results of these simulations are shown in Fig.~\ref{fig:DisplacedHamiltSim}. While there are differences in both the fidelity graphs and the displaced photon counts, the overall behavior of the different simulations is, up to different oscillations, quite comparable. Specifically, the performance of all the simulations in reaching a high fidelity, as well as the derivative of the displaced photon count, are very similar across the Hamiltonians. Since the main goals of the simulations in this work are the extraction of the derivative $d\tilde{n}_m/dt$ and the calculation of the $0.99$ fidelity times,  we deemed the final form of the Hamiltonian in Eq.~\ref{eq:convHamiltDisplacedRWAZero} appropriate for this work.
\section{MAXIMUM COOLING RATE}\label{supp:MaxCoolingRate}
The cooling rate has a ``hard'' limit which is the decay rate of the readout mode $\kappa$, in the sense that the readout mode must decay faster than it is populated. If the coupling strength to the readout mode $g_r$ becomes too strong then non-Markovian effects would reduce the effective cooling rate. This limits the cooling through the readout mode to a rate of $\kappa/4$. There is another ``soft'' limit, which comes from the fact that the qubit can only contain one excitation at a time. This second limit can be overcome by increasing the coupling rate, untill the hard limit is reached. We will show this explicitly by solving for the state of the readout mode. Since the Hamiltonian preserves the number of excitations, then the rate of photon loss from the readout mode is also the cooling rate of the memory mode. We write the Master equation corresponding to the effective Hamiltonian after applying the RWA (Eq.~\ref{eq:convHamiltDisplacedRWAZero})
\begin{equation}
\begin{split}
    \dot{\rho} &= -\mathrm{i}\left[H,\rho\right]+\kappa\mathcal{D}[\tilde{a}_r]\rho, \\
    H&=-(g_r^\star \tilde{a}_r\sigma_+g_r\tilde{a}_r^\dagger\sigma_-)-(g_m^\star \tilde{a}_m\sigma_++g_m\tilde{a}_m^\dagger\sigma_-)~,
\end{split}
\end{equation}
where $\mathcal{D}[A]\rho = A\rho A^\dagger - 1/2(A^\dagger A \rho + \rho A^\dagger A)$ is the Lindblad dissipator.
We look for a steady coherent state solution for $\langle \tilde{a}_r\rangle$, meaning that
\begin{equation}
\begin{split}
    \langle \dot{\tilde{a}}_r \rangle = \dot{\tilde{\alpha}}=-\mathrm{i}\left \langle \left[\tilde{a}_r, H \right ]\right\rangle -\frac{\kappa}{2}\langle\tilde{a}_r\rangle \\= \mathrm{i}g_r^\star\langle\sigma_-\rangle -\frac{\kappa}{2}\tilde{\alpha}_r = 0~,
    \end{split}
\end{equation}
so that we obtain
\begin{equation}
    \tilde{\alpha}_r = \frac{2\mathrm{i}g_r^\star}{\kappa}\langle\sigma_-\rangle~.
\end{equation}
The displaced frame photon loss rate is therefore limited by
\begin{equation}
    |\dot{\tilde{n}}_m|=|\tilde{\alpha}_r|^2\kappa = \frac{4|g_r|^2}{\kappa}|\langle\sigma_-\rangle|^2= \kappa|\langle\sigma_-\rangle|^2 \leq \frac{\kappa}{4}  ~,
\end{equation}
where we substituted $g_r=\kappa/2$ and bounded $|\langle\sigma_-\rangle|^2$ by
\begin{equation}
\begin{split}
    |\langle\sigma_-\rangle|^2 = \frac{1}{4}|\langle\sigma_x\rangle-\mathrm{i}\langle\sigma_y\rangle|^2\\=-\frac{1}{4}\left(|\langle\sigma_x\rangle|^2+|\langle\sigma_y\rangle|^2\right)\leq \frac{1}{4}
    \end{split}
\end{equation}

\section{ADDITIONAL SIMULATIONS AND CALCULATIONS}\label{supp:QubitParams}

In addition to the simulations presented in the article, there is value in analyzing the effect of different system parameters on the performance of RDR. While the values of $\kappa$, $\chi_m$, and $\chi_r$ (which were chosen to be identical to the values in Ref.~\cite{Milul2023SuperconductingTime}) are typical for these types of coupling setups, there are other parameters which significantly affect the efficacy of our method.
The qubit decay time constant ($T_{1}$) and dephasing time constant ($T_{2}$) were considered effectively infinite for the simulations in the article, but improvements in these characteristic time constants are an ongoing challenge in qubit fabrication today~\cite{Copetudo2024ShapingCQED}. Additionally, we took $\Omega_R$ to be $20$ MHz. Most transmons have capacitive energies $E_C$ of only around a few $100$ MHz, which allow for a Rabi frequency of a few dozen MHz at most.

Because of the challenges and limitations in achieving well performing values for these parameters, it is of import to simulate the effect of these parameters on our cooling method. We have thus simulated cooling for different values of $T_{1}$,$T_{2}$, and $\Omega_R$.

\begin{figure}
    \centering
    \includegraphics[width=0.95\linewidth,trim = 0 30 20 10,clip]{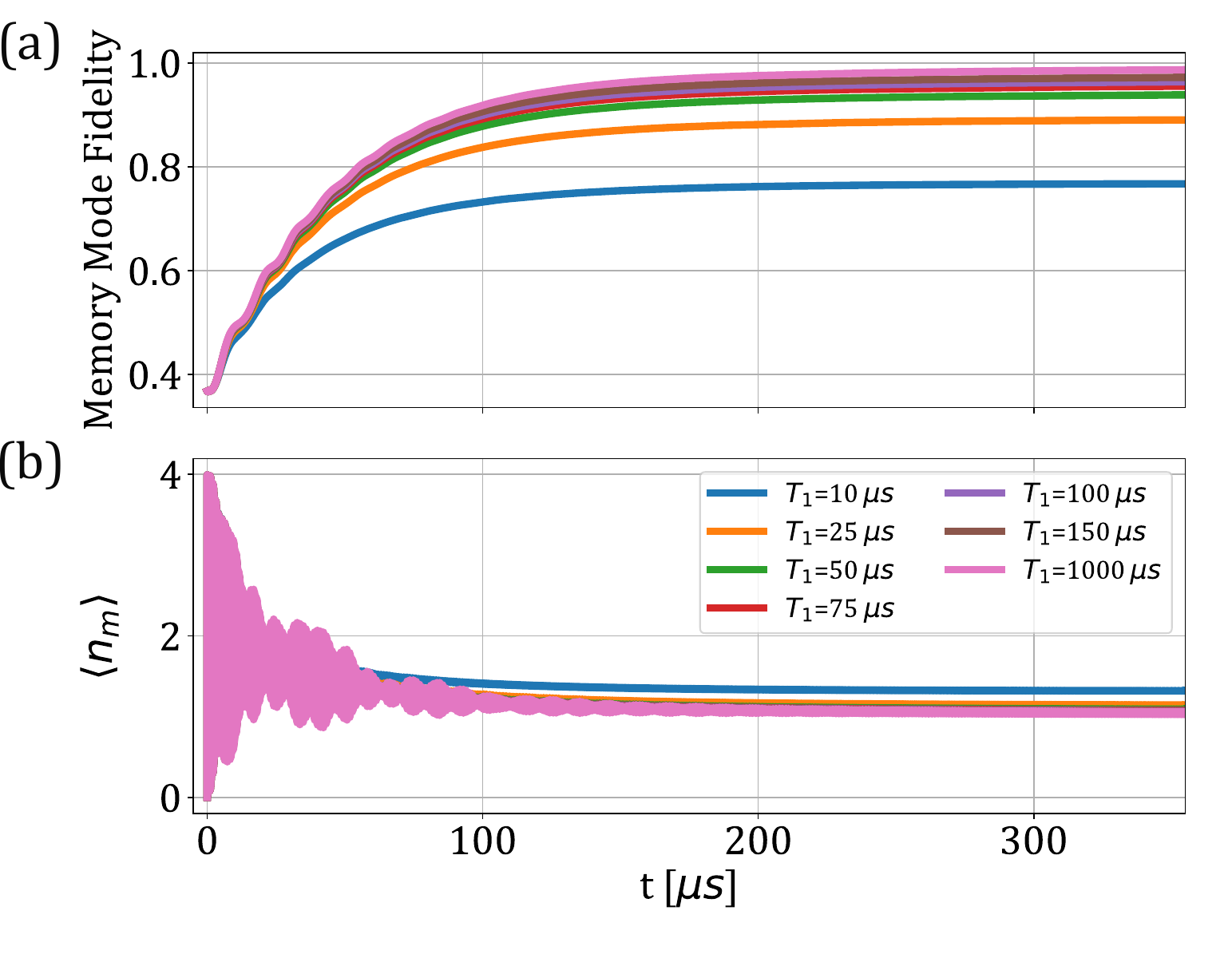}
    \caption{Simulation of the Hamiltonian in Eq.~\ref{eq:convHamilt} with final required photonic occupation $n_m=1$ for different valus of $T_1$. (a) The fidelity of the memory mode to its required final state. (b) The photonic occupation of the memory mode.}
    \label{fig:QubitT1}
\end{figure}

\begin{figure}
    \centering
    \includegraphics[width=0.95\linewidth,trim = 0 30 20 10,clip]{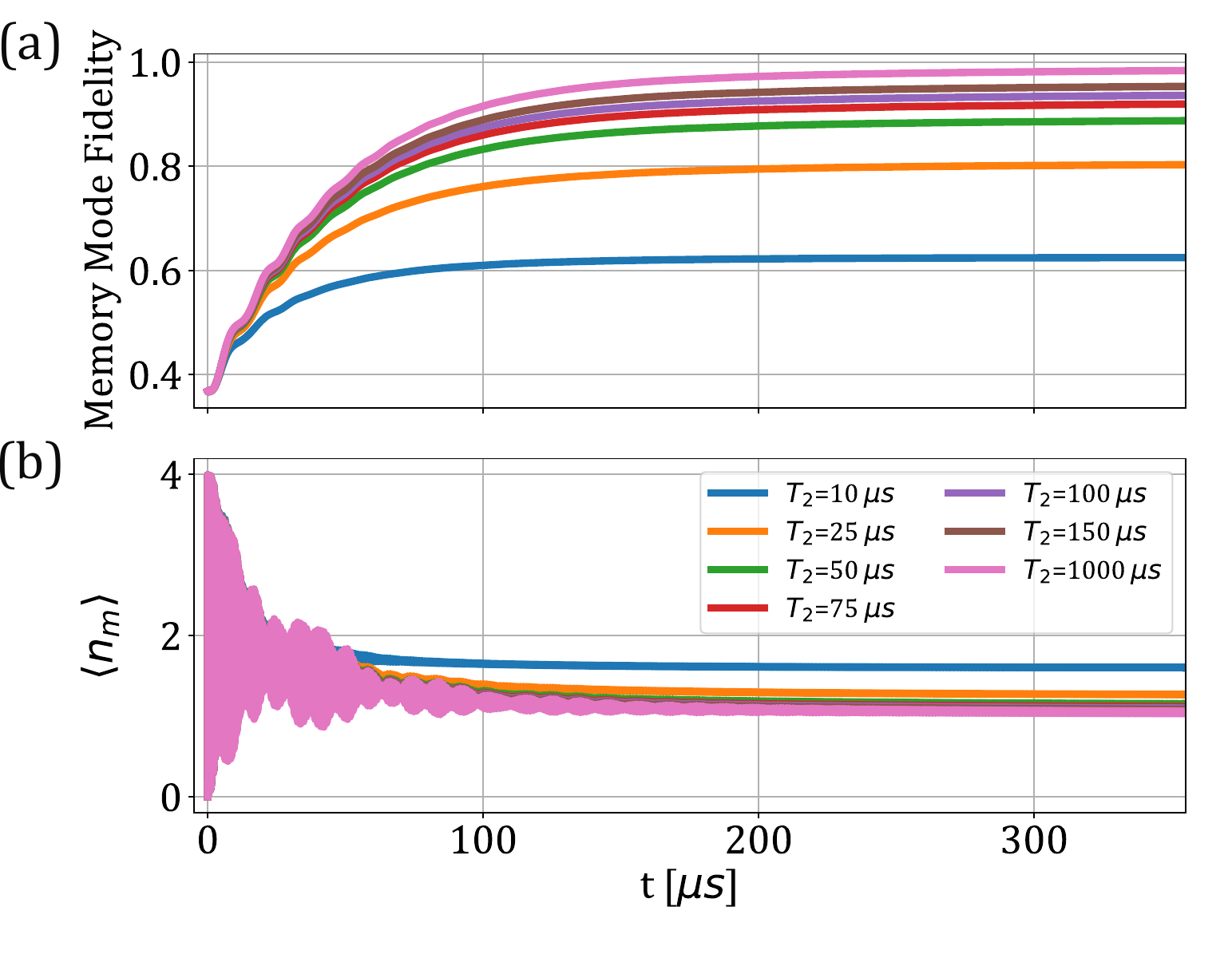}
    \caption{Simulation of the Hamiltonian in Eq.~\ref{eq:convHamilt} with final required photonic occupation $n_m=1$ for different valus of $T_2$. (a) The fidelity of the memory mode to its required final state. (b) The photonic occupation of the memory mode.}
    \label{fig:QubitT2}
\end{figure}

For $T_{1}$ and $T_{2}$ simulations, we first performed the simulation of section~\ref{sec:fullHamiltOnly} with a required final state with $n_m=1$. These simulations were performed with the full Hamiltonian of Eq.~\ref{eq:convHamilt}. Because these simulations are comparatively computationally difficult, we have analyzed the effect of $T_1$ and $T_2$ separately. The results are shown in Fig.~\ref{fig:QubitT1} and Fig.~\ref{fig:QubitT2}. Do note that since for $T_2$ simulations $T_1$ is effectively infinite, the pure dephasing time $T_\phi$ is identical to $T_2$.

It is possible to observe for different values of both $T_{1}$ and $T_{2}$ that for worsening time constants the final fidelity becomes worse, and the overall fidelity graph also has lower values (although the general shape is similar). The photon count is mostly the same for all graphs although for short time constants the photon count starts displaying a significant deviation from the target number. Additionally, it seems apparent that smaller values of $T_2$ affect the performance of our method more adversely than equivalent values of $T_1$.

\begin{figure}
    \centering
    \includegraphics[width=0.95\linewidth,trim = 0 30 20 10,clip]{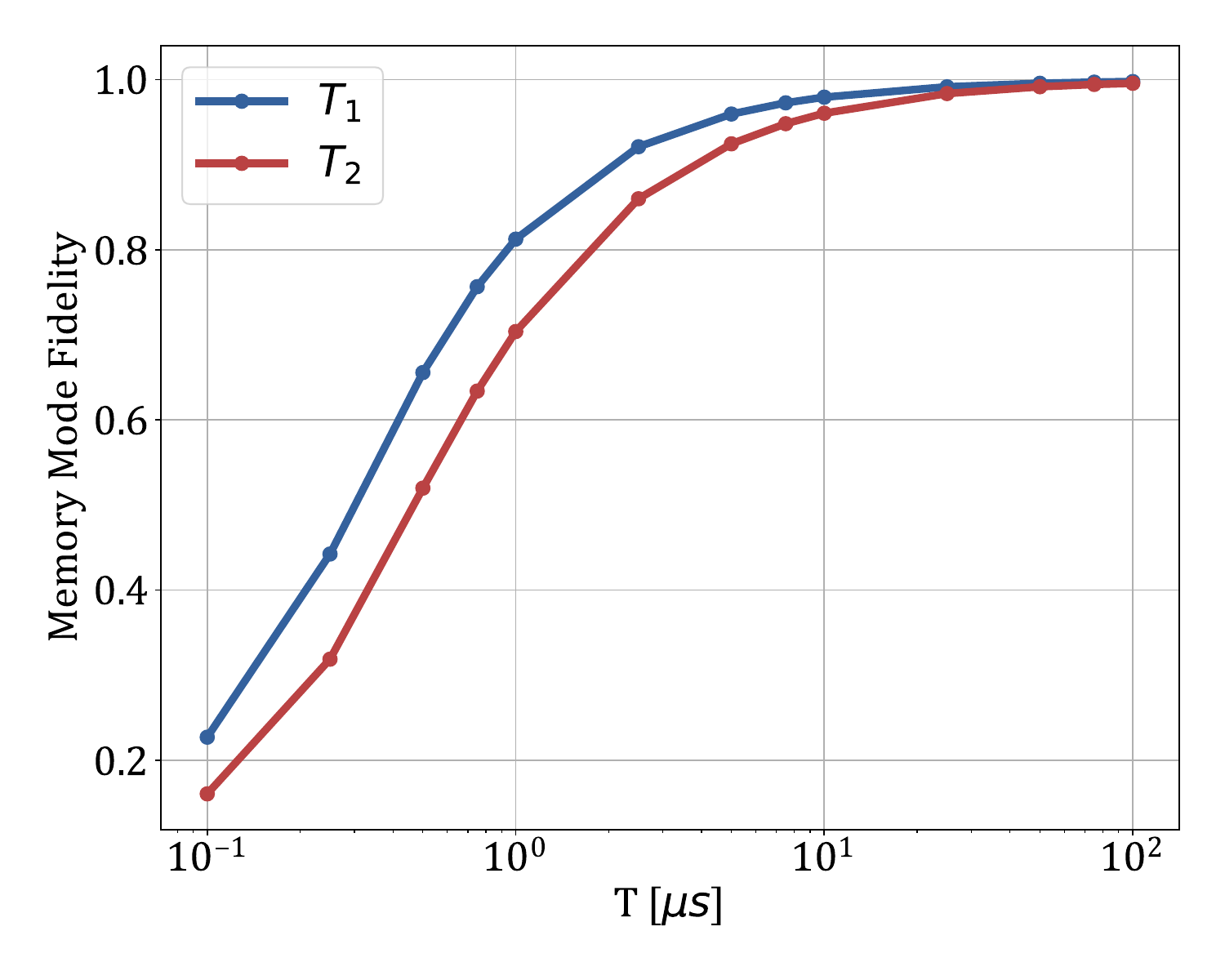}
    \caption{Calculated steady state fidelity of the memory mode as a function of $T_1$ (in blue) and $T_2$ (in red). The calculation was performed using Eq.~\ref{eq:convHamiltDisplaced} with final photonic occupation $n_m=24.8$.}
    \label{fig:QubitT1T2Steady}
\end{figure}

Despite the value of these simulations in showing the overall qualitative performance of our method with worsening $T_1$ and $T_2$, the fidelity values they show are not actually indicative of the maximum achievable fidelity using our method. This is because the decay and dephasing effects are in direct competition with our cooling drives, meaning that for weak drives (as in our $n_m=1$ simulations) the decay and dephasing rates are much ``stronger'' compared to our relatively slow cooling. However, we cannot directly calculate qubit decay in a maximum cooling rate simulation such as in section~\ref{sec:maxRate} since only the Hamiltonians of Eq.~\ref{eq:convHamiltDisplaced} or Eq.~\ref{eq:convHamilt} can correctly simulate qubit decay, which is computationally prohibitive.
Due to these constraints, we have calculated the final fidelity values for different decay and dephasing times using the QuTiP steady state solver for the Hamiltonian in Eq.~\ref{eq:convHamiltDisplaced} for final photonic occupation of $n_m=24.8$. The results of these calculations are shown in Fig.~\ref{fig:QubitT1T2Steady}. In these calculations the effects of $T_1$ and $T_2$ on the performance of our method is much smaller. This implies that even relatively simple qubits should be able to reach high fidelity values using our method. Additionally, even in these calculations, the effect of $T_2$ is slightly more significant than the effect of $T_1$ on the final fidelity of the state.

\begin{figure}
    \centering
    \includegraphics[width=0.95\linewidth,trim = 0 30 20 10,clip]{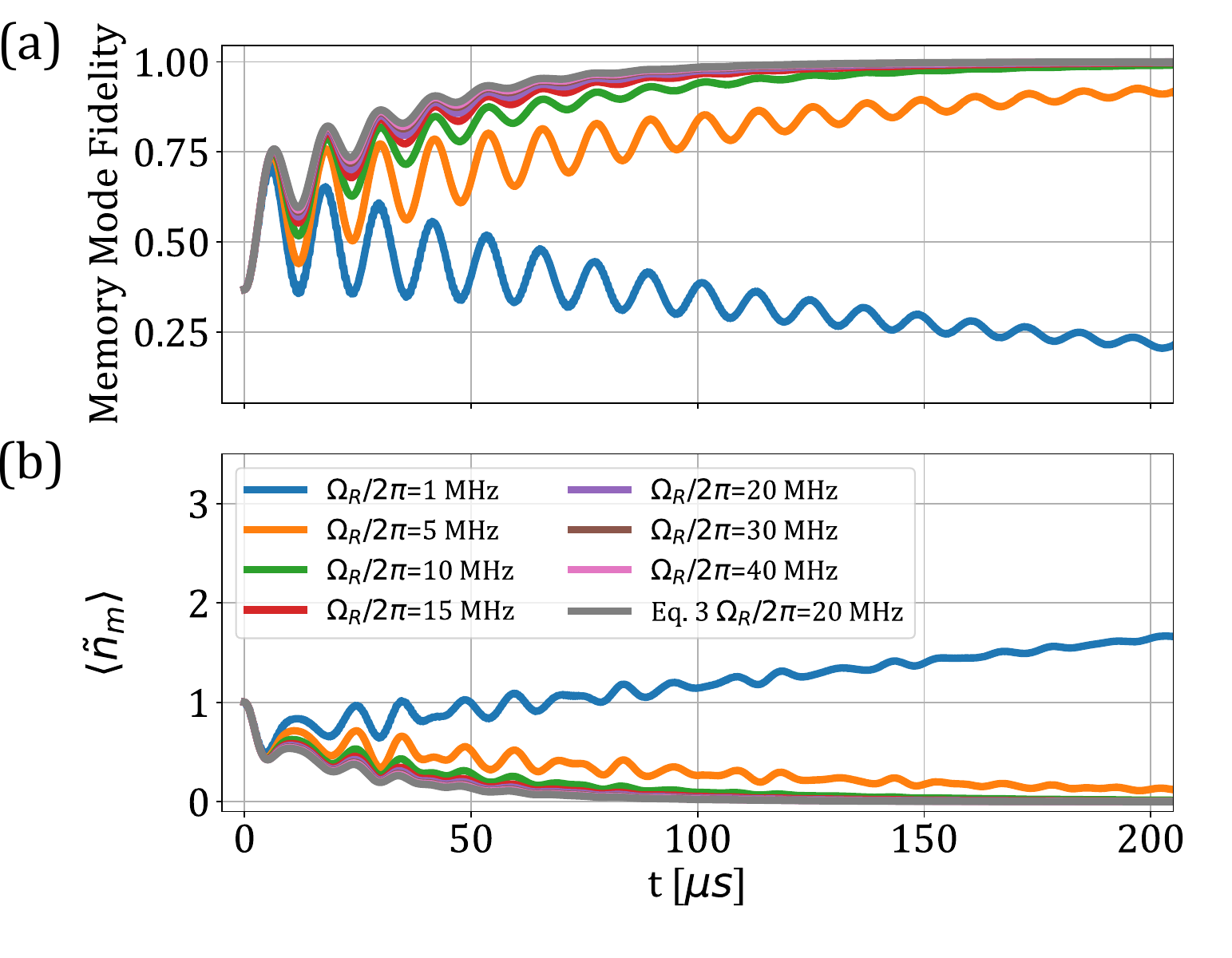}
    \caption{Simulation of the Hamiltonian in Eq.~\ref{eq:convHamiltDisplaced} with final photonic occupation $n_m=1$ for different valus of $\Omega_R$. An additional simulation using Eq.~\ref{eq:convHamiltDisplacedRWAZero} is also shown in gray. (a) The fidelity of the memory mode to its required displaced final state of $\ket{\alpha=0}$ in the displaced frame. (b) The displaced photonic occupation of the memory mode.}
    \label{fig:OmegaR}
\end{figure}

For different values of $\Omega_R$, we would need to use a different Hamiltonian than the one in Eq.~\ref{eq:convHamiltDisplacedRWAZero}, since it excludes constant energy terms (which is where the $\Omega_R$ value will come into effect). Thus, it is preferable to use the fuller Hamiltonian of Eq.~\ref{eq:convHamiltDisplaced}, since it includes $\Omega_R$ terms but is less computationally demanding compared to Eq.~\ref{eq:convHamilt}. Results of the simulations with final photonic occupation $n_m=1$ with different $\Omega_R$ values are shown in Fig.~\ref{fig:OmegaR}. They are shown together with a simulation of the simplified Eq.~\ref{eq:convHamiltDisplacedRWAZero} which does not contain $\Omega_R$ terms as part of the rotating wave approximation, as discussed in Appendix~\ref{supp:RotatingWave}.

For high enough $\Omega_R$ the simulations become virtually indistinguishable from the simulation of Eq.~\ref{eq:convHamiltDisplacedRWAZero}. This follows logically from the justification of the rotating wave approximation: The $\Omega_R$ terms must be much faster than other terms in the Hamiltonian in order to be considered approximately zero. This is due to the fact that in the rotating wave approximation these terms are assumed to rotate much faster than any change in the other terms, which is only true for high $\Omega_R$ values. 
As $\Omega_R$ becomes smaller the performance of the method becomes worse, both in the increase rate of the fidelity and the final photonic occupation. For $\Omega_R/2\pi$ values lower than $10$ MHz these divergences become dominant and it can be considered that the method ceases to perform correctly. It is of note though that compared to the simulations of the different time constants, it seems that for high enough values of $\Omega_R$, The improvements in performance are modest at most. This means that for a setup which achieves a minimum acceptable $\Omega_R$ there is little to be gained from attempting to increase $\Omega_R$.

\begin{figure}
    \centering
    \includegraphics[width=0.95\linewidth,trim = 0 30 20 10,clip]{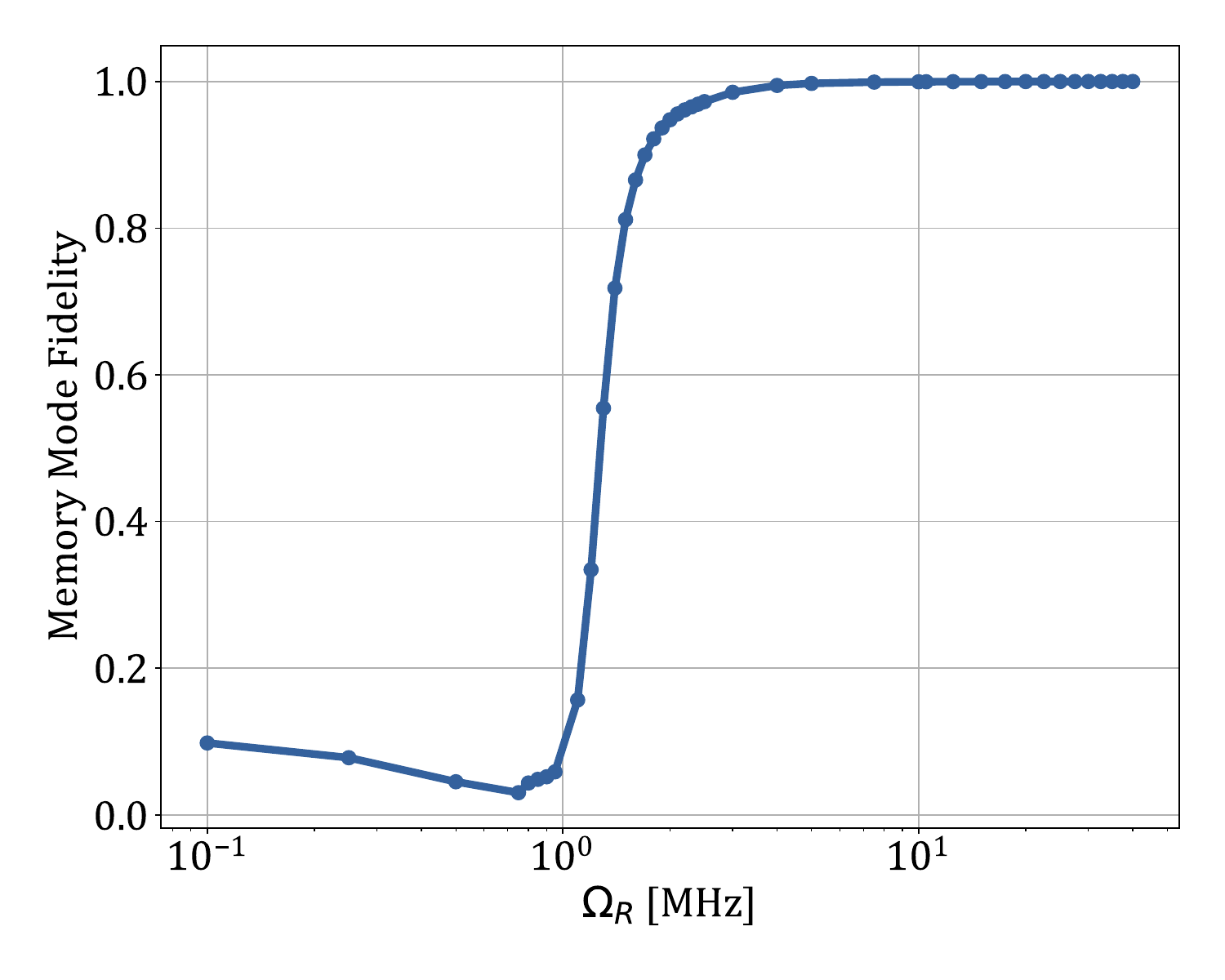}
    \caption{Calculated steady state fidelity of the memory mode as a function of $\Omega_R$. The calculation was performed using Eq.~\ref{eq:convHamiltDisplaced} with final photonic occupation $n_m=24.8$.}
    \label{fig:SteadyOmegaR}
\end{figure}

Just as for $T_1$ and for $T_2$, it would also be useful to show the steady state fidelity for different $\Omega_R$ values, using the maximum cooling rate setup used in section~\ref{sec:maxRate} with the Hamiltonian in Eq.~\ref{eq:convHamiltDisplaced}. The results of these steady state calculation are shown in Fig.~\ref{fig:SteadyOmegaR}. It is interesting to note that the effect of $\Omega_R$ is very sharp, with an increase from a low fidelity to a high fidelity beyond a certain frequency. This along with the previous simulations seems to suggest that the effect of $\Omega_R$ on the method is binary: either it is high enough for the RWA to be applicable, or it is not and the method begins to fail. It does seem that for our system parameters the $\Omega_R$ values do not need to be particularly high (around $3$ MHz) to be sufficient.

It is thus useful to summarize that $T_1$, $T_2$, and $\Omega_R$ act as thresholds to the successful operation of the RDR method, with required minimum $T_1$ and $T_2$ values of the order of $10 \mu s$ and a required minimum $\Omega_R$ value of $3$ MHz for the system parameters we have used.

\section{SIMULATION PARAMETERS}\label{supp:Sim}
We performed all QuTiP simulations as described in the article. Here we provide some additional technical details which will allow the complete reconstruction of our simulations:

 Any decoherence or dephasing times which are taken to be infinite are in fact taken to be $10^{10} ns$, which is effectively infinite in the timescales of our simulations.

 Fidelity is calculated as the probability of the simulated state in the displaced frame to be in the coherent vacuum state $\ket{\alpha=0}$.

\begin{figure}
    \centering
    \includegraphics[width=0.95\linewidth,trim = 0 30 20 10,clip]{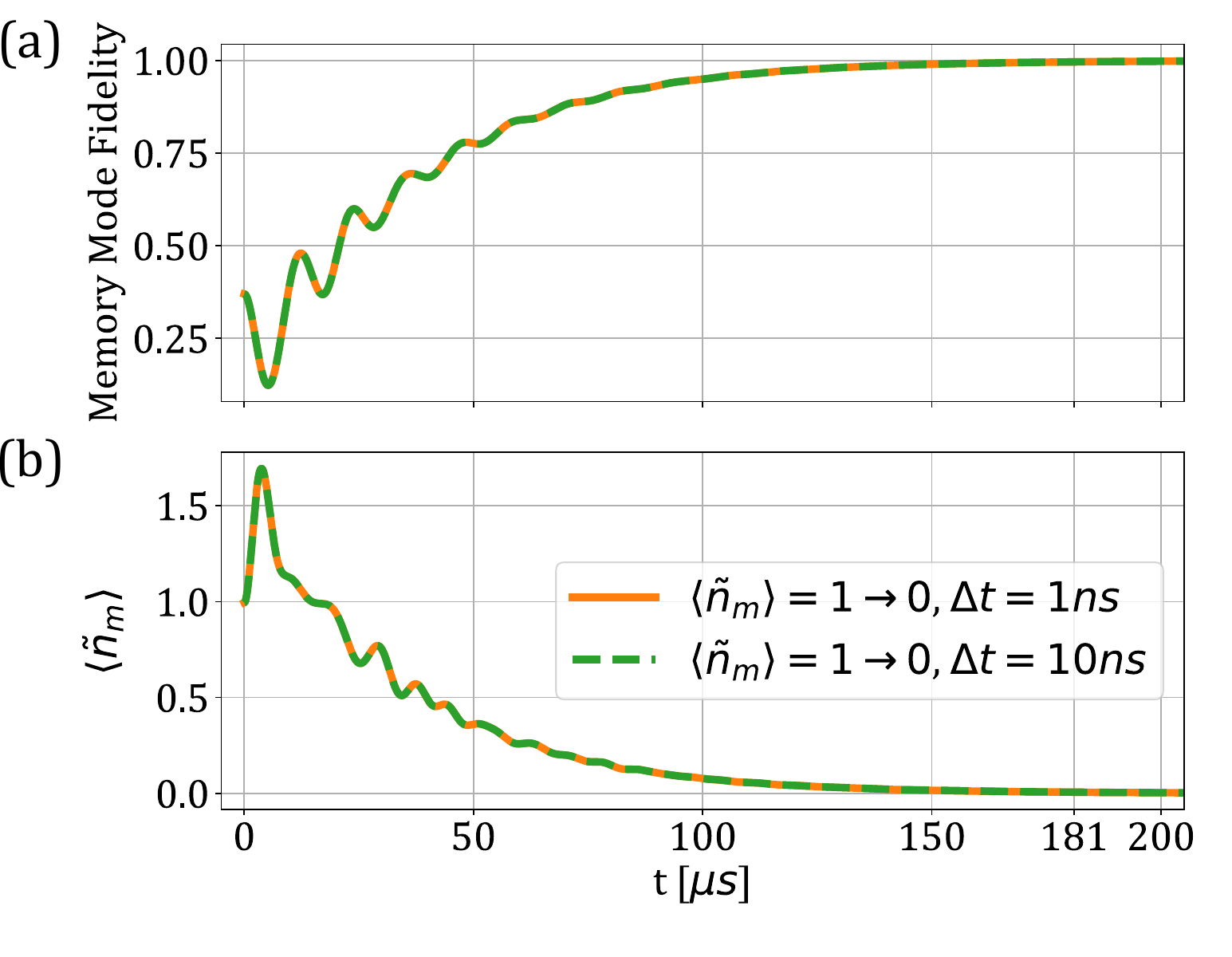}
    \caption{Simulation of the Hamiltonian in Eq.~\ref{eq:convHamiltDisplacedRWAZero} with the parameters of section~\ref{sec:fullHamiltOnly} with timestep $\Delta t=1ns$, in orange, and an increased timestep $\Delta t=10ns$, in dashed green. (a) The fidelity of the memory mode to its required displaced final state of $\ket{\alpha=0}$ in the displaced frame. (b) The displaced photonic occupation of the memory mode.}
    \label{fig:supTimestep}
\end{figure}

For the simulations using Eq.~\ref{eq:convHamilt} or Eq.~\ref{eq:convHamiltDisplaced}, the timestep was $\Delta t=1ns$. In simulations of Eq.~\ref{eq:convHamiltDisplacedRWAZero}, since we removed all high order effects, it was deemed sufficient to simulate with a longer timestep of $\Delta t=10ns$. To justify this, we compared a simulation of Eq.~\ref{eq:convHamiltDisplacedRWAZero} with the simulation parameters of the first simulation of section~\ref{sec:fullHamiltOnly} for both of these timestep values, and found that they are both functionally identical, as shown in Fig.~\ref{fig:supTimestep}.

\newpage

\bibliography{references}

\end{document}